\documentclass[aps,prx,reprint]{revtex4-1}
\usepackage{blindtext}
\usepackage{amssymb}
\usepackage{amsmath}
\usepackage{tikz}
\usepackage{graphicx}

\newtheorem{proposition}{Proposition}

\newtheorem{prop}{Proposition}\def\PRO{\begin{prop}}\def\ORP{\end{prop}}
\newtheorem{coro}{Corollary}\def\COR{\begin{coro}}\def\ROC{\end{coro}}
\newtheorem{theo}{Theorem}\def\TH{\begin{theo}}\def\HT{\end{theo}}
\def\TH{\begin{theo}}\def\HT{\end{theo}}
\newtheorem{defi}[prop]{Definition}\def\DE{\begin{defi}}\def\ED{\end{defi}}

\newtheorem{lemme}[prop]{Lemma}\def\LE{\begin{lemme}}\def\EL{\end{lemme}}
\usepackage{color}

\def\ket#1{\left| #1 \right\rangle}

\newcommand{\beq}{\begin{equation}}
\newcommand{\eeq}{\end{equation}}

\begin{document}
\title{Measurement-Device-Independent Quantum Digital Signatures}
\author{Ittoop Vergheese Puthoor$^1$}
\email{Ittoop.Puthoor@hw.ac.uk}
\author{Ryan Amiri$^1$}
\author{Petros Wallden$^2$}
\author{Marcos Curty$^3$}
\author{Erika Andersson$^1$}
\affiliation{$^1$SUPA, Institute of Photonics and Quantum Sciences, Heriot-Watt University, Edinburgh EH14 4AS, United Kingdom\\
$^2$LFCS, School of Informatics, University of Edinburgh, 10 Crichton Street, Edinburgh EH8 9AB, United Kingdom\\
$^3$EI Telecomunicaci$\acute{o}$n, Department of Signal Theory and Communications, University of Vigo, Vigo E-36310, Spain}

\begin{abstract}
Digital signatures play an important role in software distribution, modern communication and financial transactions, where it is important to detect forgery and tampering. Signatures are a cryptographic technique for validating the authenticity and integrity of messages, software, or digital documents. The security of currently used classical schemes relies on computational assumptions. Quantum digital signatures (QDS), on the other hand, provide information-theoretic security based on the laws of quantum physics. Recent work on QDS~\cite{Amiri2015, Yin2015} shows that such schemes do not require trusted quantum channels and are unconditionally secure against general coherent attacks. However, in practical QDS, just as in quantum key distribution (QKD), the detectors can be subjected to side-channel attacks, which can make the actual implementations insecure. Motivated by the idea of measurement-device-independent quantum key distribution (MDI-QKD), we present a measurement-device-independent QDS (MDI-QDS) scheme, which is secure against all detector side-channel attacks. Based on the rapid development of practical MDI-QKD, our MDI-QDS protocol could also be experimentally implemented, since it requires a similar experimental setup.
\end{abstract}

\maketitle

\section{Introduction}

Digital signatures are techniques for guaranteeing the authenticity and integrity of a message. They play a significant role for example in financial transactions, software distribution, and e-mail. Signature schemes allow a sender to exchange messages with many recipients, with the assurance that the messages cannot be forged or tampered with. In addition, signed messages are also transferable, and cannot be repudiated. Transferability means that a message, which is accepted by an honest recipient, will also be accepted by another recipient if the message is forwarded. Non-repudiation is related to transferability and means that a sender cannot successfully deny having sent a signed message. 

Classical digital signature schemes rely on public-key encryption. The security of such protocols is based on the assumed computational difficulty of inverting certain cryptographic functions. For example, an algorithm that is widely used for generating digital signatures is the Rivest-Shamir-Adleman (RSA)~\cite{Rivest1978} cryptosystem, which relies on the difficulty of factoring the product of two large prime numbers. However, if a quantum computer is built, this may threaten the security of such protocols. This is a main motivation for developing unconditionally secure  signature schemes~\cite{Swanson2011, Amiri2015a}, including quantum digital signature (QDS) schemes~\cite{Gottesman2001, Andersson2006, Clarke2012, Dunjko2014, Collins2014}. The latter are essentially quantum versions of Lamport's one-time signature scheme~\cite{Lamport1979}, and can offer information-theoretic security relying on the fundamental laws of quantum physics. 

Previous QDS schemes~\cite{Andersson2006, Clarke2012, Dunjko2014, Collins2014} improved on the seminal work in~\cite{Gottesman2001} by removing the need for quantum memory. Wallden \emph{et.al.}~\cite{Wallden2015} proposed more practical QDS schemes which could be realized using QKD~\cite{Bennett1984} components. In these QDS schemes, Alice encodes her signatures in quantum states, and sends a copy of each state to both Bob and Charlie. Bob and Charlie are only able to gain partial information on the overall signature state, due to its quantum nature. Until recently, the security analysis of all QDS schemes assumed authenticated quantum channels. In ~\cite{Amiri2015, Yin2015}, all trust assumptions on the quantum channels are removed, which is a significant improvement compared to the previous schemes. 

It is however more challenging to guarantee the security of practical implementations of QDS schemes. This is so because practical realisations do not typically conform to the requirements imposed by the theory, as real devices can behave differently from the models considered in the security proofs.
As a result, we have that any imperfection which is not accounted for might constitute a ``side channel'' which could be used by an adversary to render the QDS scheme insecure.  Here, the most critical devices are arguably the single-photon detectors~\cite{Qi2007, Lamas2007, Zhao2008, Xu2010, Lydersen2010, Weier2011, Gerhardt2011, Jouguet2013}. For example, an adversary can use detector loopholes to learn about a participant's (say Bob's) measurement results, and could then forge a message with Bob. In the context of QKD, detector side-channels can be successfully removed by means of measurement-device-independent QKD (MDI-QKD)~\cite{Lo2012}. In this approach, Alice and Bob do not perform any measurement but only send quantum signals to be measured. Thus, the advantage of MDI-QKD is that the legitimate parties need not hold a measurement device and may treat the measurement apparatus as a ``black box", which may be fully controlled by Eve. This is important as it eliminates the requirement to certify the detectors in a QKD standarization process. Therefore, the bit strings generated by Alice and Bob are free from detector side-channel attacks as they do not employ any detector. Hence, this only requires Alice and Bob to characterize the quantum states which they send through the channel. This characterization should take place in a protected environment outside the influence of the adversary, which in principle is feasible. Since the invention of MDI-QKD, such schemes have been very actively studied both theoretically~\cite{Tamaki2012, Ma2012, Xu2013, Curty2014} and experimentally~\cite{Rubenok2013, Ferreira2013, Liu2013, Tang2014, Tang2014a, Comandar2015}.

In this paper, we present a QDS protocol which eliminates all detector side-channel attacks by employing the concept of measurement-device-independence. This is desirable for actual practical use of QDS schemes. The main contribution of this work is to adapt the rigorous security proof of MDI-QKD given in~\cite{Curty2014}, taking into account finite-size effects, to the QDS protocol proposed in \cite{Amiri2015}. The resulting security proof is valid against general forging and repudiation attacks. Long-distance implementation of MDI-QKD~\cite{Rubenok2013, Ferreira2013, Liu2013, Tang2014, Tang2014a, Comandar2015} has been recently achieved, and the experimental parameters allowing for MDI-QKD could equally well allow for implementation of our QDS protocol. Hence, we envisage not just a long-distance implementation of a QDS protocol, but an implementation that is secure against detector side-channel attacks.

\section{The protocol}

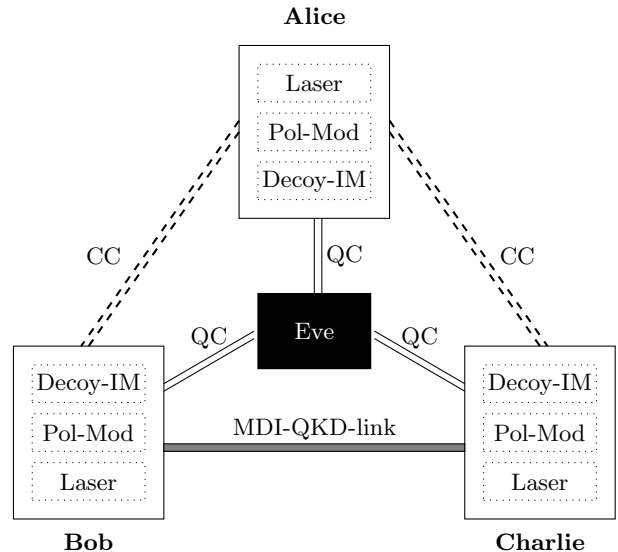
\begin{figure}[h!]
\begin{tikzpicture}
\draw[fill=white] (2,-1) rectangle (4,1.30);\draw[dotted] (2.25,-0.75) rectangle (3.75,-0.25);\draw[dotted, fill=white] (2.25,-0.10) rectangle (3.75,0.40);\draw[dotted, fill=white] (2.25,0.55) rectangle (3.75,1.05); \node at (3,-0.5){Decoy-IM};\node at (3,0.15){Pol-Mod};\node at (3,0.8){Laser};
\node at (3,1.7){$\bold{Alice}$};
\draw[fill=white] (-1,-5) rectangle (1,-2.70);
\draw[dotted] (-0.75,-4.75) rectangle (0.75,-4.25);\draw[dotted, fill=white] (-0.75,-4.10) rectangle (0.75,-3.60);\draw[dotted, fill=white] (-0.75,-3.45) rectangle (0.75,-2.95); \node at (0,-4.5){Laser};\node at (0,-3.85){Pol-Mod};\node at (0,-3.2){Decoy-IM};
\node at (0,-5.3){$\bold{Bob}$};
\draw[fill=white] (5,-5) rectangle (7,-2.70);
\draw[dotted] (5.25,-4.75) rectangle (6.75,-4.25);\draw[dotted, fill=white] (5.25,-4.10) rectangle (6.75,-3.60);\draw[dotted, fill=white] (5.25,-3.45) rectangle (6.75,-2.95); \node at (6,-4.5){Laser};\node at (6,-3.85){Pol-Mod};\node at (6,-3.2){Decoy-IM};
\node at (6,-5.3){$\bold{Charlie}$};
\draw[fill=black] (2.25,-2) rectangle (3.75,-3);\node [white] at (3,-2.5){Eve};
\draw (3,-1) -- (3,-2);\draw (3.1,-1) -- (3.1,-2);\node at (3.4,-1.5){QC};\draw (1,-3.2) -- (2.2,-2.5);\draw (1,-3.3) -- (2.2,-2.6);\node at (1.6,-2.6){QC};\draw (5,-3.2) -- (3.8,-2.5);\draw (5,-3.3) -- (3.8,-2.6);\node at (4.4,-2.6){QC};\draw[fill=gray] (1,-4) rectangle (5,-4.1);\node at (3,-3.8){MDI-QKD-link};\draw [dashed, thick] (4,0.15) -- (6,-2.7);\draw [dashed, thick] (4,0.3) -- (6.1,-2.7);\node at (5.7,-1.5){CC};\draw [dashed, thick] (2,0.15) -- (0,-2.7);\draw [dashed, thick] (2,0.3) -- (-0.1,-2.7);\node at (0.2,-1.5){CC};
\end{tikzpicture}
\caption{A schematic diagram of a setup for MDI-QDS. Alice, Bob and Charlie prepare quantum signals in different BB84 polarisation states, using a polarisation modulator (Pol-Mod). In addition, they generate decoy-states with an intensity modulator (Decoy-IM). The signals are then sent to an untrusted party Eve, who acts like a relay and is supposed to perform a Bell state measurement, which projects the incoming signals into a Bell state. The channels between Alice-Eve, Bob-Eve and Charlie-Eve are quantum channels (QC). Eve performs the measurement separately for the pairs Alice-Bob and Alice-Charlie. Bob and Charlie share a MDI-QKD link (grey channel), which can be used to transmit classical messages in full secrecy. The pairs Alice-Bob and Alice-Charlie have pairwise authenticated classical channels (CC) indicated as dashed lines, through which they can communicate their basis settings for the different key positions.}
\label{fig:mdiqds}
\end{figure}

We outline our protocol for three parties, with a sender, Alice, and two recipients Bob and Charlie. The set-up for MDI-QDS is illustrated in Fig.~\ref{fig:mdiqds}. We assume that between Alice and Bob, and between Alice and Charlie, there exist authenticated classical channels. There is no need for ``direct" quantum channels between Alice and Bob, between Alice and Charlie, nor between Bob and Charlie. Each party has an untrusted and imperfect quantum channel with the relay (Eve). 
Bob and Charlie share a MDI-QKD link, which can be used to transmit classical messages in full secrecy. This is separately indicated in the figure, but could also be realised with Eve as relay. Any classical secret communication channel between Bob and Charlie would in fact suffice in place of this MDI-QKD link. We will describe the procedure for signing a one-bit message. For signing longer messages, the procedure can be suitably iterated, meaning that the signature length scales linearly with message length.

Alice, Bob and Charlie each use a laser source to generate quantum signals that are diagonal in the Fock basis. Sources producing such signals include attenuated laser diodes emitting phase-randomised weak coherent pulses (WCPs), triggered spontaneous parametric down-conversion sources and practical single-photon sources. The scheme makes use of a measurement-device-independent key generating protocol (MDI-KGP), performed in pairs separately by Alice-Bob and Alice-Charlie; see Section~\ref{sec:mdikgp} for more details. The purpose of such an MDI-KGP scheme is to use the noisy untrusted quantum channels to generate two correlated bit strings, one for each participant in an MDI-KGP.
The noise level is defined in terms of the relative Hamming distance between these strings. When the noise level is below a tolerated value, the relative Hamming distance between the respective strings of the participants is smaller than the relative Hamming distance between any string that an eavesdropper could produce, and the participant's string. 

The QDS scheme above is related to the one proposed in ~\cite{Amiri2015}, with a difference in the KGP. It comprises of two stages, a distribution stage, where all quantum communication takes place, and a messaging stage, which can occur much later, and where only classical communication is used.

\subsection{Distribution stage}

\noindent(1) For each possible future message $m$=0 or 1, Alice uses the MDI-KGP to generate four different correlated bit strings, $A^B_0, A^B_1, A^C_0, A^C_1$, each one of length $L$. The superscript denotes the participant with whom Alice performed the MDI-KGP, and the subscript represents the future message, which is to be decided later by her. Bob holds the strings $K^B_0, K^B_1$ and Charlie holds the strings $K^C_0, K^C_1$. Because of the KGP, it will be guaranteed that $A^B_0$ contains fewer mismatches with $K^B_0$ than does any string produced by an eavesdropper, and similarly for the other pairs of strings. Alice's signature for the future message $m$ will be $Sig_m = (A^B_m, A^C_m)$.  The fact that only Alice knows all signatures for a message $m$ protects the protocol against forging. \\

\noindent (2) For each future message, Bob and Charlie symmetrize their keys. This is done by each of them choosing at random half of the bit values in their keys ($K^B_m, K^C_m$) and sending these bit values (as well as the corresponding positions) to the other participant using their secret classical channel. This will ensure that Alice cannot make Bob and Charlie disagree on the validity of a signature, if a message is forwarded from Bob to Charlie or vice versa in the messaging stage. If Bob (or Charlie) chooses to forward an element of $K^B_m$ (or $K^C_m$) in the distribution stage to Charlie (or Bob), he will not, if he is honest, further use it to check the validity of a signature. Bob and Charlie will only use the bits they did not forward, and those received from the other participant. This is not strictly necessary, but simplifies the analysis of repudiation by a dishonest Alice in that from Alice's point of view, the probabilities are equal for Bob and Charlie to check a particular key bit. We denote their symmetrized keys by $S^B_m$ and $S^C_m$, with the superscript indicating whether the key is held by Bob or Charlie. Bob (and Charlie) keep a record of whether an element in $S^B_m$ ($S^C_m$) came directly from Alice or whether it was forwarded to him by Charlie (or Bob).

Each of the symmetrized strings held by Bob and Charlie now contains half of $K^B_m$ and half of $K^C_m$. For each future possible message $m$, Bob and Charlie each have a bit string of length $L$. Alice has no information on whether it is Bob's $S^B_m$ or Charlie's $S^C_m$ that contains a particular element of the string $(K^B_m, K^C_m)$, which is of length $2L$. This protects against repudiation. Bob has access to all of $K^B_m$ and half of $K^C_m$. He does not know the other half of $K^C_m$ which Charlie chose to keep. This protects the protocol against forging by Bob (and similarly against forging by Charlie).

\subsection{Messaging stage}
\noindent (1) To send a signed one-bit message $m$, Alice sends $(m, Sig_m)$ to the desired recipient (say Bob).\\

\noindent (2) Bob checks whether $(m, Sig_m)$ matches his $S^B_m$, and records the number of mismatches he finds. He separately checks the part of his key received directly from Alice and the part of the key received from Charlie. If there are fewer than $s_a(L/2)$ mismatches in both halves of the key, where $s_a<1/2$ is a small threshold determined by the observed experimental parameters (see Appendix ~\ref{App:AppendixD} for more details) and the desired security level of the protocol, then Bob accepts the message.\\ 

\noindent (3) To forward the message to Charlie, Bob forwards the pair $(m, Sig_m)$ that he received from Alice.\\

\noindent (4) Charlie tests for mismatches in a similar way, but using a different threshold in order to protect against repudiation by Alice. He accepts the forwarded message if the number of mismatches in both halves of his key is below $s_v(L/2)$ where $s_v$ is another threshold, with $0 < s_a < s_v < 1/2$. An important and necessary feature of unconditionally secure signature schemes~\cite{Swanson2011, Arrazola2015} is that the recipients have to use different thresholds or acceptance criteria for messages received directly from the sender and for forwarded messages.

\section{Measurement-device-independent key generation protocol}
\label{sec:mdikgp}

MDI-QKD protocols~\cite{Lo2012, Curty2014, Xu2015a} are schemes that remove all detector side-channel attacks. This is very important when we consider detector loopholes in conventional QKD implementations~\cite{Qi2007, Jouguet2013}. Similarly, the key generation protocol, which is part of the QDS scheme we are describing, can be made measurement-device-independent. Essentially, Alice and Bob (or Alice and Charlie) only perform the quantum part of the MDI-QKD scheme to generate raw different keys (the $A^{B}_{m}$ and $K^{B}_{m}$ described above) with imperfectly correlated and not completely secret bit strings. That is, Alice and Bob do not perform error correction and privacy amplification. This is sufficient for quantum signatures, since it is the number of mismatches with the recipient's key that matters for the signature protocol; perfectly correlated, perfectly secret strings are not necessary. The aim is to show that $\Lambda(A^{B}_{m}, K^{B}_{m}) < \Lambda(E_{guess}, K^{B}_{m})$ except with negligible probability, where $\Lambda(x,y)$ is the Hamming distance between $x$ and $y$, and $E_{guess}$ is Eve's attempt at guessing $K^{B}_{m}$. It can also be possible that the adversary Eve is Charlie (for the KGP performed between Alice and Bob, and for the KGP performed by Alice and Charlie, Eve could be Bob). The security of the signature protocol is proved in Sec.~\ref{Sec:Security}.

The underlying MDI-QKD protocol, upon which the KGP is built, is the decoy-state BB84 protocol using phase-randomized WCPs considered in~\cite{Lo2012}. We follow the steps of the protocol in~\cite{Curty2014}, using the $Z$ basis for key generation, but do not proceed with error correction and privacy amplification.  

The different steps of the MDI-KGP are as follows.

\noindent (1) {\bf State preparation:} Alice and Bob repeat the first two steps of the protocol for $i = 1,...,N$ until the conditions in the Sifting stage are met. For each $i$, Alice chooses an intensity $a \in \{a_s, a_{d_{1}}, a_{d_{2}} \}$, a basis $\alpha \in \{Z,X\}$, and a random bit $r \in \{0,1\}$ with probability $p_{a,\alpha}/2$. Here $a_s$ ($a_{d_{j}}$ where $j\in\{1,2\}$) is the intensity of the signal (decoy) states. Next, she generates a quantum signal (e.g, a phase-randomized WCP) of intensity $a$ prepared in the basis state of $\alpha$ given by $r$. Similarly, Bob does the same. Alice and Bob then send their states to Eve via the quantum channel.

\noindent (2) {\bf Measurement:} If Eve is honest, she makes a Bell state measurement of the signals she has received. Whether Eve is honest or not, she informs Alice and Bob through a public channel of whether or not her measurement was successful. If successful, she declares the Bell state that is obtained.

\noindent (3) {\bf Sifting:} If Eve reports a successful result, Alice and Bob communicate through an authenticated channel their intensity and basis settings. For each Bell state $k$, we define two groups of sets: $Z_{k}^{a,b}$ and $X_{k}^{a,b}$. $Z_{k}^{a,b}$ is a set that identifies signals where Eve declares a Bell state $k$ and Alice and Bob have selected the intensities $a$ and $b$ and the basis $Z$. Similarly, $X_{k}^{a,b}$ is a set that identifies signals where Eve declares a Bell state $k$ and Alice and Bob have selected the intensities $a$ and $b$ and the basis $X$. The protocol is repeated until $|Z_{k}^{a,b}| \geq N_{k}^{a,b}$ and $|X_{k}^{a,b}| \geq M_{k}^{a,b}$ $\forall a, b, k$~\cite{footnote1}. After this, Bob flips part of his bits to correctly correlate them with those of Alice. 
This is shown in Table~\ref{tab:sifting}.
\begin{table}[h!]
    \begin{tabular}{ | c | c | c | c | c |}
    \hline
    & \multicolumn{4}{| c |} {Bell state reported by Eve} \\ \hline
     Alice's \& Bob's basis & $\ket{\psi^{-}}$ & $\ket{\psi^{+}}$ & $\ket{\phi^{-}}$ & $\ket{\phi^{+}}$  \\ \hline
     $Z$ basis & Bit flip & Bit flip &  -- & -- \\ \hline
     $X$ basis & Bit flip & -- &  Bit flip & -- \\ \hline
    \end{tabular}
  \caption{Processing of data in the sifting stage. The Bell states are defined as $\ket{\psi^{-}} = \frac{1}{\sqrt{2}}(\ket{HV} - \ket{VH}), \ket{\psi^{+}} = \frac{1}{\sqrt{2}}(\ket{HV} + \ket{VH}), \ket{\phi^{+}} = \frac{1}{\sqrt{2}}(\ket{HH} + \ket{VV})$ and $\ket{\phi^{-}} = \frac{1}{\sqrt{2}}(\ket{HH} - \ket{VV})$.}
  \label{tab:sifting}
\end{table}

\noindent (4) {\bf Parameter Estimation:} Alice and Bob use $n_{k}$ random bits from $Z_{k}^{a_{s},b_{s}}$ to form the code bit strings $\mathcal{Z}_{k}$ and $\mathcal{Z'}_{k}$, respectively. The remaining $R_{k}$ bits from $Z_{k}^{a_{s},b_{s}}$ are used to compute the error rate $E_{k}^{a_{s},b_{s}} = \frac{1}{R_{k}}\sum_{l}r_{l}\oplus r_{l'}$ where $r_{l}$ and $r_{l'}$ are Alice's and Bob's bits respectively. The  bit string of length $R_{k}$ is used to estimate the correlation between Alice and Bob's strings generated from the $Z$ basis, after which they are discarded. If $E_{k}^{a_{s},b_{s}} > E_{tol}~\forall k$, then Alice and Bob abort the protocol. 
If $E_{k}^{a_{s},b_{s}} \leq E_{tol}$, Alice and Bob use $Z_{k}^{a,b}$ and $X_{k}^{a,b}$ to estimate $n_{k,0}, n_{k,1}$ and $e_{k,1}$. The parameter $n_{k,0}$ is a lower bound for the number of bits in $\mathcal{Z'}_{k,\text{keep}}$ where Bob sent a vacuum state. $\mathcal{Z'}_{k,\text{keep}}$ is the part of $\mathcal{Z'}_{k}$ which he chooses to keep with himself while he forwards the other remaining part, $\mathcal{Z'}_{k,\text{forward}}$, to Charlie during the key symmetrization process. That is, $|\mathcal{Z'}_{k,\text{keep}}|=|\mathcal{Z'}_{k,\text{forward}}|=n_{k}/2$. In a similar way, $n_{k,1}$ is a lower bound for the number of bits in $\mathcal{Z'}_{k,\text{keep}}$ where Alice and Bob sent a single-photon state. $e_{k,1}$ is an upper bound for the single-photon phase error rate. If $e_{k,1} \geq e_{tol}$, the code bit strings $\mathcal{Z}_{k}$ and $\mathcal{Z'}_{k}$ are discarded, and the protocol is aborted only if $e_{k,1} \geq e_{tol}$ $\forall k$.

\begin{figure}[h!]
\begin{tikzpicture}
\node at (4.5,3.5){$\bold{Eve}$};
\draw[ultra thick] (1.5,-0.5) rectangle (7.5,4); 
\draw [-] (4.5,-0.25) -- (4.5,0.75);\draw [-] (4.5,-0.25) -- (4,0.25);\draw [-] (4,0.25) -- (4.5,0.75);
\draw [-] (4.5,-0.25) -- (5,0.25);\draw [-] (5,0.25) -- (4.5,0.75);\node at (5.25,0.25){BS};
\draw [->] (4.25,0.5) -- (3.5,1.25);\draw [->] (4.75,0.5) -- (5.5,1.25);
\draw [->] (6,-1.25) -- (4.75,0);\draw [<-] (4.25,0) -- (3,-1.25);
\draw [thick] (3.25,1) -- (3.25,2);\draw [thick] (3.25,1) -- (3.75,1.5);\draw [thick] (3.75,1.5) -- (3.25,2);
\draw [thick] (3.25,1) -- (2.75,1.5);\draw [thick] (2.75,1.5) -- (3.25,2);\node at (2.4,1.5){PBS};
\draw [->] (3,1.75) -- (2.5,2.25);\draw [->] (3.5,1.75) -- (4,2.25);
\draw [thick] (5.75,1) -- (5.75,2);\draw [thick] (5.75,1) -- (5.25,1.5);\draw [thick] (5.25,1.5) -- (5.75,2);
\draw [thick] (5.75,1) -- (6.25,1.5);\draw [thick] (6.25,1.5) -- (5.75,2);\node at (6.7,1.5){PBS};
\draw [->] (5.5,1.75) -- (5,2.25);\draw [->] (6,1.75) -- (6.5,2.25);
\draw [-] (2.3,2.05) -- (2.7, 2.45);\draw [gray] (2.7,2.45) arc [radius=0.4, start angle=90, end angle=180];\node at (2.4,2.7){$D_{1H}$};
\draw [-] (4.2,2.05) -- (3.8, 2.45);\draw [gray] (4.2,2.05) arc [radius=0.4, start angle=0, end angle=90];\node at (4,2.7){$D_{1V}$};
\draw [-] (4.8,2.05) -- (5.2, 2.45);\draw [gray] (5.2,2.45) arc [radius=0.4, start angle=90, end angle=180];\node at (5,2.7){$D_{2V}$};
\draw [-] (6.7,2.05) -- (6.3, 2.45);\draw [gray] (6.7,2.05) arc [radius=0.4, start angle=0, end angle=90];\node at (6.5,2.7){$D_{2H}$};
\draw[fill=white] (2,-1.25) rectangle (4,-2);\node at (3,-1.6){Alice};
\draw[fill=white] (5,-1.25) rectangle (7,-2);\node at (6,-1.6){Bob/Charlie};
\end{tikzpicture}
\caption{A schematic diagram of Eve's measurement device. The combination of polarising beam splitters (PBSs) and a 50:50 beam splitter (BS) projects the incoming signals from Alice and Bob/Charlie into horizontal (H) and vertical (V) polarisation states. A joint click on the single-photon detectors $D_{1H}$ and $D_{2V}$, or $D_{1V}$ and $D_{2H}$, represents a projection into the Bell state $\ket{\psi^{-}}$, while a joint click in $D_{1H}$ and $D_{1V}$, or $D_{2V}$ and $D_{2H}$, indicates a projection into the Bell state $\ket{\psi^{+}}$.}
\label{fig:mdiqds2}
\end{figure}
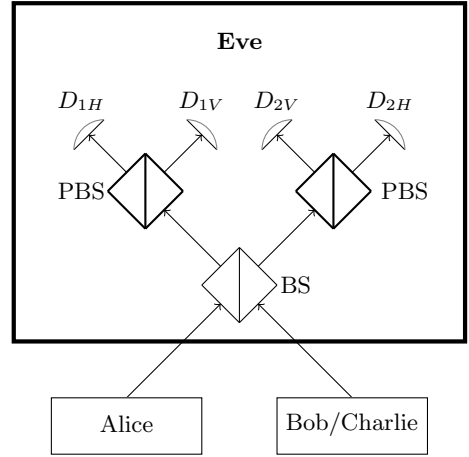
 
We will assume that Eve implements her Bell state measurement using linear optics. The measurement setup is illustrated in Fig.~\ref{fig:mdiqds2}; it is able to identify two of the four Bell states. Alice and Bob choose $\mathcal{Z}_{k}$ and $\mathcal{Z'}_{k}$ as their respective secret keys $A^{B}_{m}$ and $K^{B}_{m}$ of length $L$ (where $L = n_{k}$), for which they obtained the smallest phase error rate $e_{k,1}$. 
Here, we will consider a finite number of states that are sent and measured, where Eve is allowed to perform general coherent attacks. 

Our strategy is to find Eve's information in terms of the smooth min-entropy~\cite{Tomamichel2011}, and then use it to bound the probability that she can make a signature declaration making fewer errors than a certain value. We begin by finding Eve's smooth min-entropy on Bob's bit string $\mathcal{Z'}_{k,\text{keep}}$, by following the same strategy as in~\cite{Amiri2015}. In spite of the fact that the KGP is built on MDI-QKD, the security analysis for the MDI-KGP does not follow directly from the security of the MDI-QKD protocol. One reason is that the goal of an adversary in the signature protocol is different from that of an eavesdropper in MDI-QKD. For the signature protocol, what matters is the number of mismatches with a recipient's key; for QKD, what matters is the information an eavesdropper can hold about a key. These are related but not identical. 

Previous work~\cite{Amiri2015} followed~\cite{Lim2014} to find Eve's smooth min-entropy in a similar way as for decoy-state QKD. Another important difference from QKD is that in the signature protocol, Bob effectively gives the extra information $\mathcal{Z'}_{k,\text{forward}}$ to Eve (with respect to forging with Bob, Charlie can be ``Eve"). In a similar way, let us denote the classical random variables $R_{k}$ and $\Theta$ as the information gained by Eve from parameter estimation and basis declarations for all the pulses sent by Alice and Bob, respectively. Since Bob, if he is honest, does not use $\mathcal{Z'}_{k,\text{forward}}$, this could be treated as the part of the string $R_{k}$ that is sacrificed for parameter estimation, as explained in~\cite{Tomamichel2015}. We combine all of Eve's information into one quantum system living in the Hilbert space $\mathcal{H}_{E}$. This comprises the space containing Eve's ancilla quantum system following her general attack, $\mathcal{H}_{E'}$, as well as the spaces containing the states encoding the strings $R_{k}, \Theta$ and $\mathcal{Z'}_{k,\text{forward}}$.
Then, according to~\cite{Curty2014}, Eve's smooth min-entropy, which quantifies the average probability that she guesses $\mathcal{Z'}_{k,\text{keep}}$ within a certain threshold using the optimal strategy with access to $E_{k}$, is given by
\beq \label{eq:main}
\begin{split}
H_{\text{min}}^{\varepsilon_{k}}(\mathcal{Z'}_{k,\text{keep}} | E_{k})_{\rho} \geq \hspace{4cm} \\n_{k,0} + n_{k,1}\left[1 - h(e_{k,1})\right]- 2\log_{2}\frac{2}{\varepsilon'_{k}\hat{\varepsilon}_{k}},
\end{split}
\eeq
where $\varepsilon_k \geq \varepsilon'_k + {\hat \varepsilon}_k$ and $\rho$ is the state shared by Eve and the part of the key that Bob kept and did not forward. We are interested in a regime where the first two terms on the RHS of equation~\eqref{eq:main} are much larger than the $\log_{2}$ term as $\varepsilon'_k$ and $\hat{\varepsilon}_k$ are typically of the order say $10^{-5} -10^{-10}$. Therefore, we arrive at the following approximation of equation~\eqref{eq:main}:
\begin{eqnarray} \label{eq:mains}
H_{\text{min}}^{\varepsilon_{k}}(\mathcal{Z'}_{k,\text{keep}} | E_{k})_{\rho} &\gtrapprox & n_{k,0} + n_{k,1}\left[1 - h(e_{k,1})\right].
\end{eqnarray}
Appendix~\ref{App:AppendixA} provides a brief analysis of the estimation of the parameters $n_{k,0}$, $n_{k,1}$, and $e_{k,1}$, and Appendix~\ref{App:AppendixB} briefly describes the steps involved to obtain equation~\eqref{eq:main}. 

Note that equation~\eqref{eq:mains} is similar to equation (1) obtained in~\cite{Amiri2015}. The next task is to bound the number of errors that Eve is likely to make when guessing Bob's key, given the bound on her smooth min-entropy. For this, we use Proposition $1$ in~\cite{Amiri2015} and follow the same argumentation.

\begin{proposition}\cite{Amiri2015}
If Bob and Eve share the state $\rho$
then, for any eavesdropping strategy, Eve's average probability of making at most $r$ mistakes when guessing $\mathcal{Z'}_{k,\text{keep}}$ can be upper bounded as
\begin{equation} \label{eq:result2}
\langle p_r \rangle \leq \sum^r_{m=0} \binom{\frac{n_{k}}{2}}{m} 2^{-H^{\varepsilon_{k}}_{\text{min}}(\mathcal{Z'}_{k,\text{keep}}|E_{k})_{\rho}} + \varepsilon_{k}.
\end{equation}
\end{proposition}
The proof of this proposition follows the lines introduced in Appendix B of~\cite{Amiri2015}. 
For large $n_{k}$, it can be shown from Markov's inequality that equation~\eqref{eq:result2} implies
\beq \label{eq:forgeprob1}
Pr(\text{Eve makes fewer than r errors}) := p_r \leq g,
\eeq
except with probability at most
\beq \label{eq:forgeprob2}
 p_F := \frac{1}{g} \left(2^{-\frac{n_{k}}{2}\left\{c_{k,0} + c_{k,1} [1-h(e_{k,1})] - h(2r/n_{k})\right\}} + {\varepsilon_{k}}\right),
\eeq
where $c_{k,i} := {2n_{k,i}}/{n_{k}}$ is the lower bound on the count rate for the $Z$ basis pulses containing $i$ photons. Therefore, we arrive at the condition that determines whether or not Eve is able to make fewer than $r$ errors with non-negligible probability, given as
\beq
c_{k,0} + c_{k,1} [1-h(e_{k,1})] - h(2r/n_{k}) > 0.
\eeq
If the condition holds, then $n_{k}$ can be increased to make Eve's probability of making fewer than $r$ errors arbitrarily small. We define $p_{E}$ by the equation
\beq \label{eq:EveError}
c_{k,0} + c_{k,1} [1-h(e_{k,1})] - h(p_{E}) = 0.
\eeq
The meaning of this is that $p_{E}$ is the minimum rate at which Eve can make errors for the code string associated with the Bell state $k$ (except with negligible probability $p_F$). Suppose the error rate on the $Z$ basis measurements between Alice and Bob is upper bounded as $\overline{E}_{k}$. As long as $p_{E}>\overline{E}_k$, there exists a choice of parameters and a sufficiently large signature length which makes the protocol secure. This means that MDI-QDS is possible as long as
\beq \label{eq:coh_range}
c_{k,0} + c_{k,1} [1-h(e_{k,1})] - h(\overline{E}_k) > 0.
\eeq 

\section{Security analysis}
\label{Sec:Security}

We will now prove the security of the signature protocol, i.e. the robustness (probability of an honest run aborting), security against forging (probability that a recipient generates a signature, not originating from Alice, that is accepted as authentic) and repudiation (or transferability) (probability that Alice generates a signature that is accepted by Bob but then when forwarded, is rejected by Charlie). In what follows we assume that Alice-Bob and Alice-Charlie have each used the MDI-KGP to generate bit strings of length $L = n_{k}$, to use in the QDS protocol described above.

\emph{(a) Robustness}. Bob rejects a signed message if the $\frac{n_{k}}{2}$ bits received from either Alice or Charlie have a mismatch rate higher than $s_a$ with Alice's signature. We note that Alice and Bob use a random sample, $R_{k}$ bits from $Z_{k}^{a_{s},b_{s}}$, to obtain the error rate $E_{k}^{a_{s},b_{s}}$. This implies that the error rate $\overline{E}_{k}^{a_{s},b_{s}}$ between the strings ($\mathcal{Z}_{k,\text{keep}}$ and $\mathcal{Z'}_{k,\text{keep}}$) generated using the $Z$ basis satisfies the  inequality~\cite{Serfling1974}
\begin{equation} \label{eq:actualerrorrate}
\overline{E}_{k}^{a_{s},b_{s}} \geq E_{k}^{a_{s},b_{s}} + \mu\left(\frac{n_{k}}{2},R_{k},\varepsilon_{PE}\right) ,
\end{equation}
where 
\begin{equation}\label{eq:fp}
\mu\left(\frac{n_{k}}{2},R_{k},\varepsilon_{PE}\right) =  \sqrt{\frac{(\frac{n_{k}}{2} - R_{k} + 1)\ln({\frac{1}{\varepsilon_{PE}}})}{R_{k}n_{k}}}.
\end{equation}
This means that the upper bound which we obtain from equation~\eqref{eq:actualerrorrate} on the error rate between Alice's and Bob's strings is true except with a very small probability $\varepsilon_{PE}$, and this probability can be fixed as small as desired. For any fixed value of the function $\mu$, the failure probability decays exponentially fast in the parameter $R_{k}$. 
Then we set $\overline{E}_{k} := max\{\overline{E}_{k,B}, \overline{E}_{k,C}\}$, where $\overline{E}_{k,B}$ and $\overline{E}_{k,C}$ refer to the upper bound obtained in equation~\eqref{eq:actualerrorrate} for the cases Alice-Bob and Alice-Charlie, and  we choose $s_{a}$ such that $s_a > \overline{E}_{k}$. We have that
the probability that Bob will find an error rate higher than $s_a$ is bounded by
\begin{equation} \label{eq:honab}
Pr(\text{Honest Abort}) \leq  2\varepsilon_{PE},
\end{equation}\\
where the factor of 2 accounts for the fact that the abort can be due to either the states received from Alice or the states received from Charlie.

\emph{(b) Security against repudiation.} Successful repudiation by Alice means, in the three-party scenario, that she makes Bob accept a declaration $(m, Sig_m)$ that was sent to him by her, while Charlie rejects the same declaration when Bob forwards it to him (or similarly for a message forwarded from Charlie to Bob). Intuitively, security against repudiation follows because of the symmetrisation performed by Bob and Charlie using the secret classical channel. Even if Alice knows and can control the error rates between $A^B_m$, $A^C_m$ and $K^B_m$, $K^C_m$, she cannot control whether the errors end up with Bob or Charlie. After symmetrisation the keys $S^B_m$ and $S^C_m$ will each have the same expected number of errors. To repudiate, one key must contain significantly more errors than the other. Using results from~\cite{Amiri2015}, we obtain
\begin{equation} \label{eq:rep}
Pr(\text{Repudiation}) \leq 2\exp\left[-\frac{1}{4}(s_v-s_a)^2n_{k}\right].
\end{equation}
For a formal proof, please see Appendix~\ref{App:AppendixC}. Note that the probability of repudiation decays exponentially as the length $n_k$ of the signature increases.

\emph{(c) Security against forging.} It is easier for either Bob or Charlie to forge than it is for any other external party. Therefore, we will consider forging by an internal party. In order to forge a message, Bob must give a declaration $(m, Sig_m)$ to Charlie that has fewer than $s_v {n_{k}}/{2}$ mismatches with the (to Bob) unknown  half of $S^C_m$ sent directly from Alice to Charlie, and also fewer than $s_v {n_{k}}/{2}$ mismatches with the half he himself forwarded to Charlie. An adversarial Bob will obviously be able to meet the threshold on the part he forwarded to Charlie. We therefore consider only the unknown half that Charlie received directly from Alice. We have that the maximum rate at which Alice will make errors with Charlie's key is given by $\overline{E}_{k}$. From Eq. \eqref{eq:EveError}, we also know the minimum rate at which Bob will make errors with the code string associated with the Bell state $k$ of Charlie's key; we have denoted this by $p_{E}$. Assuming \eqref{eq:coh_range} holds, we choose $s_v$ such that $\overline{E}_{k}< s_v < p_{E}$. In this case, Charlie will likely accept a legitimate signature sent by Alice, since the upper bound on their error rate, $\overline{E}_{k}$, is less than the threshold $s_v$. On the other hand, Charlie will likely reject any dishonest signature declaration by Bob, since the probability of Bob finding a signature with an error rate smaller than $s_v$ is restricted by \eqref{eq:forgeprob1} as
\beq \label{eq:forge1}
Pr(\text{Bob makes fewer than $s_{v}n_{k}/2$ errors}) := p_r \leq g
\eeq
except with probability at most $p_F$ given by \eqref{eq:forgeprob2}.
If the estimation of the parameter $\overline{E}_{k}$ fails, which can happen with probability $\varepsilon_{PE}$, we will assume for simplicity that Bob is able to successfully forge with certainty. In a similar way as in~\cite{Amiri2015}, we are then able to bound Bob's probability of successfully forging as
\begin{equation} \label{eq:forge}
Pr(\text{Forge}) \leq p_F + g + \varepsilon_{PE} + \varepsilon_{k,0} + \varepsilon_{k,1} + \varepsilon_{k,e}.
\end{equation}
This equation is valid for any choice of parameters ($g, \varepsilon_{PE}, \varepsilon_{k,0}, \varepsilon_{k,1}, \varepsilon_{k,e}$) greater than $0$. Thereby, Bob's probability to forge can be made arbitrarily small by increasing $n_{k}$. The addition of $\varepsilon_{PE}$ accounts for the probability that the upper bound on $\overline{E}_{k}$ is incorrect and $\varepsilon_{k,0}, \varepsilon_{k,1}$ and $\varepsilon_{k,e}$ are the error probabilities associated with the estimation of $n_{k,0}, n_{k,1}$ and $e_{k,1}$ respectively (see Appendix \ref{App:AppendixA}). 

\section{Comparison to MDI-QKD}

\begin{table*}[t]
   \centering
    \begin{tabular}{ | c | c | c | c | c |}
    \hline
    Detectors & $\eta_D (\%)$ & $Y_0 (\times 10^{-6})$ & $N_{sig}(\times 10^{12})$ & $t_{r}$(min)\\ \hline
   Standard single-photon detectors~\cite{Ursin2007} & 14.5 & 6.02 & 5.58 & 93 \\ \hline
   InGaAs avalanche photodiodes detectors (APD)~\cite{Comandar2015} & 30 & 130 & 1.8 & 30 \\ \hline
   InGaAs/InP APD~\cite{Comandar2015b} & 55 & 500 & 0.87 & 14.5 \\ \hline
   Superconducting nanowire single-photon detectors (SNSPDs)~\cite{Marsili2013} & 93 & 1 & 0.098 & 1.6 \\ \hline
    \end{tabular}
  \caption{Raw key generation times for various detectors that could be used in a MDI-QDS protocol for a distance of 50 km and a security threshold of $10^{-5}$. The parameters $\eta_D (\%), Y_0$ and $N_{sig}$ denote respectively the detection efficiency, dark count rate of Eve's detectors, and the number of signals that Bob/Charlie sends to Alice during their KGPs. $t_{r}$ is the time taken to generate the raw key and to estimate $t_r$ we assume a source with a pulse rate of 1 GHz. 
  }
  \label{tab:detectors}
\end{table*}

\begin{table*}[t]
   \centering
    \begin{tabular}{ | c | c | c |}
    \hline
    Detectors & $N_{sig}(\times 10^{12})$ & $t_{r}$(min)\\ \hline
   Standard single-photon detectors~\cite{Ursin2007} & 10.5 & 175 \\ \hline
   InGaAs APD~\cite{Comandar2015} & 3.35 & 55.83 \\ \hline
   InGaAs/InP APD~\cite{Comandar2015b} & 1.63 & 27.1 \\ \hline
   SNSPDs~\cite{Marsili2013} & 0.18 & 3 \\ \hline
    \end{tabular}
  \caption{Raw key generation times for a distance of 50 km with a security threshold of $10^{-10}$. For the definition of the different parameters, see the caption of Table~\ref{tab:detectors}.}
  \label{tab:detectors1}
\end{table*}

According to~\cite{Curty2014}, in MDI-QKD the length $l_{k}$ of the secret bit string associated to the Bell state $k$ is given by
\begin{eqnarray} \label{eq:mdiqkd}
l_{k} &\leq& n_{k,0} + n_{k,1}[1 - h(e_{k,1})] - leak_{EC,k} -\log_{2}\frac{8}{\epsilon_{cor}} \nonumber\\ 
&&- 2 \log_{2}\frac{2}{\varepsilon'_{k}\hat{\varepsilon_{k}}} - 2 \log_{2}\frac{1}{2\varepsilon_{k,PA}},
\end{eqnarray}
if the protocol is $\epsilon_{sec}$-secret, with $\epsilon_{sec} = \sum_{k}\epsilon_{k,sec}$ and $\epsilon_{k,sec} = 2(\varepsilon'_{k} + 2\varepsilon_{k,e} + \hat{\varepsilon}_{k}) + \varepsilon_{k,b} + \varepsilon_{k,0} + \varepsilon_{k,1} + \varepsilon_{k,PA}$. Here $\varepsilon_{k,PA}$ is the failure probability of privacy amplification, and the term $leak_{EC,k}$ is the information that is revealed by Alice in the error correction step. The meaning of the remaining epsilons can be found in~\cite{Curty2014}. The correctness of the protocol is guaranteed by the error correction step, and we say that the protocol is $\epsilon_{cor}$-correct if the probability that Alice's and Bob's bit strings are not identical is not greater than $\epsilon_{cor}$. In the asymptotic limit of very large data blocks, one can neglect certain terms that reduce the secret key length and thereby equation~\eqref{eq:mdiqkd} can be rewritten as
\begin{equation} \label{eq:Finite_qkd}
l_{k} \approx n_{k}\{c_{k,0} + c_{k,1} [1-h(e_{k,1})]\} - leak_{EC,k}.
\end{equation}
Here, $c_{k,i} := {n_{k,i}}/n_{k}$ increase 
the secret key rate, while $n_{k}c_{k,1}h(e_{k,1})$ and $leak_{EC,k}$ reduce it. These parameters depend on the sifted key length $n_{k}$~\cite{Curty2014}. $leak_{EC,k} = n_{k}\zeta h(\overline{E}_{k}^{a_{s},b_{s}})$, where $\zeta$ is referred to as the leakage parameter, which depends on the value of $n_{k}$, and $h(.)$ denotes the binary Shannon entropy. $\zeta$ is assumed to be 1.16 in~\cite{Curty2014} but can generally be in the range 1.1 - 1.2, and when $n_{k} < 10^{5}$ the parameter $\zeta$ may be greater than 1.16. Therefore, for a sifted key length $\frac{n_{k}}{2}$, equation~\eqref{eq:Finite_qkd} can be written as
\begin{equation} \label{eq:Finite_qkd1}
l_{k} \approx \frac{n_{k}}{2}\{c_{k,0} + c_{k,1} [1-h(e_{k,1})] - \zeta h(\overline{E}_k)\}.
\end{equation}
In a similar way as in~\cite{Amiri2015}, when we compare equations \eqref{eq:coh_range} and \eqref{eq:Finite_qkd1}, we find that there are Alice-Bob and Alice-Charlie quantum channels for which quantum signatures are possible and yet practical MDI-QKD is not, since the error threshold is less strict for the quantum channels used to perform the KGP in the signature protocol.

\section{Discussion} 

In this section, we analyse the number of quantum transmissions necessary to sign a message with a security level of the order of $10^{-5}$ and $10^{-10}$ respectively. If the security level of the protocol is of the order of, say, $10^{-5}$, then this means that the probabilities of honest abort, forging and repudiation are all less than $10^{-5}$.

Using realistic experimental quantities, we estimate that a signature length of $n_k =  8.9\times 10^{6}$ (for each of the possible single bit messages $0$ and $1$) can be used to securely sign a single bit message, sent over a distance of 50 km. Essentially, it would require Bob or Charlie to transmit approximately $N_{sig} = 5.58\times10^{12}$ quantum states (per bit to be signed) to Alice during their KGPs (for full details see Appendix~\ref{App:AppendixD}). With a source with a pulse rate of 1GHz, we can calculate that it would take approximately 93 minutes to generate a raw key when the experiment uses standard single-photon detectors with detection efficiency ($\eta_D$) of 14.5\%. This is for a security level of the order of $10^{-5}$. By using detectors with higher detection efficiency we can improve the time of generating a raw key ($t_r$) since sending a smaller number of signals ($N_{sig}$) is then required to sign a single-bit message. 

Table~\ref{tab:detectors} shows the raw key generation times for various detectors that could be used in the protocol. We find that the most advanced superconducting nanowire single-photon detectors (SNSPDs) having $93\%$ efficiency~\cite{Marsili2013} would only require Bob or Charlie to send $6.4\times10^{10}$ signals to perform the protocol with a secure threshold of the order of $10^{-5}$. This would require just above a minute to generate the raw key. In order to improve the security threshold of the protocol (say $10^{-10}$), Bob or Charlie would need to send a higher number of signals compared to the previous case. Table~\ref{tab:detectors1} shows the raw key generation times and the number of signals that are required to send for the protocol to be secure for a threshold of the order of $10^{-10}$. 

The protocol is secure to the order of $10^{-10}$ for a distance of 50 km, which in comparison is an improvement over the previous scheme~\cite{Amiri2015} having with a security threshold of $10^{-4}$. 
The simulation results demonstrate that even with practical signals (for example, phase-randomised WCPs) and a finite size of data (say $10^{11}$ to $10^{14}$ signals) it is possible to perform secure MDI-QDS (with security threshold $10^{-10}$) over long distances (up to about 150 km). Since the experimental platform for the implementation of MDI-QKD can also be used for MDI-QDS with slight modifications, in particular in the post-processing of measurement results, we expect MDI-QDS could be widely used in practical QDS systems in the near future. 

\section{Conclusion}

In summary, we have presented a MDI-QDS protocol and proven it unconditionally secure against general attacks. It improves on previous quantum signature protocols by removing all detector side-channel attacks. This is essentially achieved by adapting the rigorous security proof of MDI-QKD given in~\cite{Curty2014}, taking into account finite size effects, to the QDS protocol proposed in \cite{Amiri2015} and we have presented that the resulting security proof is valid against general forging and repudiation attacks. 

\begin{acknowledgments} The authors would like to thank Marco Lucamarini for discussions. This work was supported by the UK Engineering and Physical Sciences Research Council (EPSRC) under EP/M013472/1. R. A. acknowledges the support of the EPSRC CM-CDT. M.C. gratefully acknowledges support from the Galician Regional Government (program ``Ayudas para proyectos de investigacion desarrollados por investigadores emergentes'' EM2014/033, and consolidation of Research Units: AtlantTIC), the Spanish Ministry of Economy and Competitiveness (MINECO), the Fondo Europeo de Desarrollo Regional (FEDER) through grant TEC2014-54898-R.
\end{acknowledgments}

\appendix
\section{Estimation of relevant parameters}\label{App:AppendixA}

In this Appendix we briefly discuss the estimation of the parameter $n_{k,0}$. This is a two-step process. First, we calculate a lower bound for the number of indices in $Z_{k}^{a_{s},b_{s}}$ where Bob sent a vacuum state. This lower bound is denoted $m_{k,0}$. Second, we compute $n_{k,0}$ from $m_{k,0}$ using the Serfling inequality for random sampling without replacement~\cite{Serfling1974}. The other parameters, $n_{k,1}$ and $e_{k,1}$, are also estimated using a similar approach. A detailed explanation is provided in the supplementary notes of \cite{Curty2014}. 

We assume that Alice and Bob use two decoy states each and the photon-number distribution of their signals is Poissonian. That is, $a \in A = \{a_{s},a_{d_{1}},a_{d_{2}}\}$, with $a_{s} > a_{d_{1}} > a_{d_{2}}$,  $b \in B = \{b_{s},b_{d_{1}},b_{d_{2}}\}$, with $b_{s} > b_{d_{1}} > b_{d_{2}}$, and the probability that Alice (Bob) sends an $n$-photon ($m$-photon) signal when she (he) selects the intensity $a$ ($b$) is given by $p_{n|a} = e^{-a}a^{n}/n!$ ($p_{m|b} = e^{-b}b^{m}/m!$). 

Let $S_{k,nm}$ denote the number of signals sent by Alice and Bob with $n$ and $m$ photons respectively, when they select the basis $Z$ and Eve declares the Bell state $k$. Now, for each combination of values $n$ and $m$, the signal and decoy states provide a random sample of the population of all signals containing $n$ and $m$ photons respectively. Therefore, one can apply the standard large deviation theory technique, in particular a multiplicative form of the Chernoff bound~\cite{Curty2014}. Then, if 
$$\left(2\varepsilon^{-1}_{a,b}\right)^{1/\mu_{k,L}^{a,b}} \leq \exp\left[3/(4\sqrt{2})]^{2}\right],$$ 
and 
$$\left(\hat{\varepsilon}^{-1}_{a,b}\right)^{1/\mu_{k,L}^{a,b}} \leq \exp\left(1/3\right),$$
with the parameter $\mu_{k,L}^{a,b}$ given by
\begin{equation}\label{eq:mu}
\mu_{k,L}^{a,b} = |Z_{k}^{a,b}| - \sqrt{\sum_{a,b}|Z_{k}^{a,b}| /2 \ln(1/\epsilon_{a,b})},
\end{equation}
this implies that
\begin{equation}\label{eq:Zab}
|Z_{k}^{a,b}| = \sum_{n,m} p_{a,b|nm,Z}S_{k,nm} + \delta_{a,b},
\end{equation}
except with error probability $\gamma_{a,b} = \epsilon_{a,b} + \varepsilon_{a,b} + \hat{\varepsilon}_{a,b}$.
Here, $p_{a,b|nm,Z}$ refers to the conditional probability that Alice and Bob have selected the intensity settings $a$ and $b$ respectively, given that their signals contain $n$ and $m$ photons respectively, prepared in the $Z$ basis. The parameter $\delta_{a,b} \in \left[-\Delta_{a,b},\hat{\Delta}_{a,b}\right]$ with $\Delta_{a,b} = g(|Z_{k}^{a,b}|,\varepsilon_{a,b}^{4}/16)$ and ${\hat{\Delta}_{a,b}} = g(|Z_{k}^{a,b}|,\hat{\varepsilon}_{a,b}^{3/2})$, and the function $g(x,y) = \sqrt{2x\ln(y^{-1})}$.

By using similar arguments, the quantity $m_{k,0}$ can be written as
\begin{equation}\label{eq:mk}
m_{k,0} = \sum_{n} p_{a_{s},b_{s}|n0,Z}S_{k,n0} - \Delta_{0},
\end{equation}
except with error probability $\varepsilon_{0}$, where $\Delta_{0} = g\left(\sum_{n} p_{a_{s},b_{s}|n0,Z}S_{k,n0}, \varepsilon_{0}\right)$. To obtain a lower bound for $m_{k,0}$, one can minimise equation~\eqref{eq:mk} given the linear constraints imposed by equation~\eqref{eq:Zab} $\forall a,b$. This is solved both analytically and numerically in the supplementary notes of~\cite{Curty2014}. Then using Serfling inequality \cite{Serfling1974}, we find
\begin{equation}\label{eq:nko}
n_{k,0} = \max \Bigg\{ \Bigg \lfloor \frac{n_{k}}{2} \frac{m_{k,0}}{|Z_{k}^{a_{s},b_{s}}|} - \frac{n_{k}}{2}\Lambda(|Z_{k}^{a_{s},b_{s}}|,\frac{n_{k}}{2}, \varepsilon''_{k,0})\Bigg \rfloor, 0 \Bigg\},
\end{equation}
except with error probability
\begin{equation}\label{eq:enko}
\varepsilon_{k,0} \leq \varepsilon'_{k,0} + \varepsilon''_{k,0},
\end{equation}
where $\varepsilon'_{k,0}\leq \varepsilon_{0} + \sum_{a,b}\gamma_{a,b}$ corresponds to the total error probability in the estimation of $m_{k,0}$ and the function $\Lambda(x,y,z)$ is defined as $\Lambda(x,y,z) = \sqrt{(x-y+1)\ln(z^{-1})/(2xy)}$. 

A similar approach is followed to estimate $n_{k,1}$ and $e_{k,1}$ with associated error probabilities $\varepsilon_{k,1}$ and $\varepsilon_{k,e}$ respectively. We obtain
\begin{equation}\label{eq:nk1}
n_{k,1} = \max \Bigg\{ \Bigg \lfloor \frac{n_{k}}{2} \frac{m_{k,1}}{|Z_{k}^{a_{s},b_{s}}|} -  \frac{n_{k}}{2}\Lambda(|Z_{k}^{a_{s},b_{s}}|, \frac{n_{k}}{2}, \varepsilon''_{k,1})\Bigg \rfloor, 0 \Bigg\},
\end{equation}
except with error probability
\begin{equation}\label{eq:enko}
\varepsilon_{k,1} \leq \varepsilon'_{k,1} + \varepsilon''_{k,1},
\end{equation}
where $\varepsilon'_{k,1} \leq \varepsilon_{1} + \sum_{a,b}\gamma_{a,b}$. Here, $m_{k,1} = p_{a_{s},b_{s}|11,Z}S_{k,11} - \Delta_{1}$, except with error probability $\varepsilon_{1}$ where the parameter $\Delta_{1} = g(p_{a_{s},b_{s}|11,Z}S_{k,11}, \varepsilon_{1})$. Finally, the parameter $e_{k,1}$ is given as
\begin{equation}\label{eq:ek1}
\begin{split}
e_{k,1} = \min \Bigg\{ \Bigg \lceil n_{k,1}\Bigg(\frac{\overline{e}_{k,1}}{\overline{n}_{k,1}}\Bigg) + (n_{k,1}+\overline{n}_{k,1})\\ \times \Upsilon(n_{k,1},\overline{n}_{k,1}, \varepsilon'''_{k,e})\Bigg \rceil, n_{k,1}  \Bigg\},
\end{split}
\end{equation}
except with error probability
\begin{equation}\label{eq:enko}
\varepsilon_{k,e} \leq \varepsilon'_{k,e} + \varepsilon''_{k,e} + \varepsilon'''_{k,e},
\end{equation}
where the function $\Upsilon (x,y,z)$ is defined as $\Upsilon (x,y,z) = \sqrt{(x+1)\ln(z^{-1})/(2y(x+y))}$. The quantity $\overline{n}_{k,1}$ is a lower bound for the number of signals where Alice and Bob send a single-photon state prepared in the $X$ basis and where Eve declares the Bell state $k$, $\overline{e}_{k,1}$ is an upper bound for the total number of errors in these signals, and $\varepsilon'_{k,e}$ and $\varepsilon''_{k,e}$ represent, respectively, their associated error probabilities.
For more details about how to calculate these parameters, please see~\cite{Curty2014}. 

We have, therefore, that the error probability associated with the estimation of the different parameters is given by $\varepsilon_{PE}+\varepsilon_{k,0} + \varepsilon_{k,1} + \varepsilon_{k,e}$, with $\varepsilon_{PE}$ given by equation~\eqref{eq:actualerrorrate}. 

\section{Eve's smooth-min entropy}\label{App:AppendixB}

The goal of this Appendix is to derive equation~\eqref{eq:chain 3}. The analysis follows the procedure introduced in~\cite{Curty2014}. For this, let $H^{\varepsilon_{k}}_{\text{min}}(\mathcal{Z'}_{k,\text{keep}}|E_{k})$ denote the smooth min-entropy which quantifies the average probability that the adversary guesses $\mathcal{Z'}_{k,\text{keep}}$ correctly using the optimal strategy with access to $E_{k}$. Now the bits of $\mathcal{Z'}_{k,\text{keep}}$ can be distributed among three different strings, $\mathcal{Z}^{'0}_{k,\text{keep}}, \mathcal{Z}^{'1}_{k,\text{keep}}$ and $\mathcal{Z}^{'\text{rest}}_{k,\text{keep}}$. The first string contains bits where Bob sent a vacuum state, the second where Alice and Bob sent a single-photon state, and $\mathcal{Z}^{'\text{rest}}_{k,\text{keep}}$ contains the rest of bits. 
Using the result of chain rule of entropies~\cite{Vitanov2013}, we obtain
\begin{equation} \label{eq:chain 1}
\begin{split}
H^{\varepsilon_{k}}_{\text{min}}(\mathcal{Z}'_{k,\text{keep}}|E_{k})
 \geq \hspace{5.5cm} \\ H_{\text{min}}^{\epsilon'_{k} + 2\epsilon''_{k} + (\hat{\epsilon_{k}} + 2\hat{\epsilon}'_{k} + \hat{\epsilon}_{k}'')}(\mathcal{Z}_{k,\text{keep}}^{'0}\mathcal{Z}_{k,\text{keep}}^{'1}\mathcal{Z}_{k,\text{keep}}^{'\text{rest}} | E_{k}) \\
\geq n_{k,0} + H_{\text{min}}^{\epsilon''_{k}}(\mathcal{Z}_{k,\text{keep}}^{'1} | \mathcal{Z}_{k,\text{keep}}^{'0} \mathcal{Z}_{k,\text{keep}}^{'\text{rest}}E_{k}) - 2\log_{2}\frac{2}{\epsilon'_{k}\hat{\epsilon}_{k}},
\end{split}
\end{equation} 
where $\varepsilon_{k} = \epsilon'_{k} + 2\epsilon''_{k} + (\hat{\epsilon_{k}} + 2\hat{\epsilon}'_{k} + \hat{\epsilon}_{k}'')$. Here, it is taken into consideration that $H_{\text{min}}^{\hat{\epsilon}'_{k}}(\mathcal{Z}_{k,\text{keep}}^{'\text{rest}} | \mathcal{Z}_{k,\text{keep}}^{'0}E_{k}) \geq 0$, and $H_{\text{min}}^{\hat{\epsilon}''_{k}}(\mathcal{Z}_{k,\text{keep}}^{'0} | E_{k}) \geq H_{\text{min}}^{0}(\mathcal{Z}_{k,\text{keep}}^{'0} | E_{k}) = H_{\text{min}}(\mathcal{Z}_{k,\text{keep}}^{'0}) = n_{k,0}$. The final part arises as the vacuum states contain no information about their bit values, which are uniformly distributed.
In order to get the lower bound for the term $H_{\text{min}}^{\epsilon''_{k}}(\mathcal{Z}_{k, \text{keep}}^{'1} | \mathcal{Z}_{k,\text{keep}}^{'0}\mathcal{Z}_{k,\text{keep}}^{'\text{rest}}E_{k})$, it is considered that Alice and Bob prepare perfect BB84 states. Then, this quantity can be written in terms of the smooth max-entropy between them, which is directly bounded by the strength of the correlations~\cite{Tomamichel2012}. From the entropy uncertainty relation~\cite{Tomamichel2011}, we obtain
\begin{eqnarray} \label{eq:chain 2}
H_{\text{min}}^{\epsilon''_{k}}(\mathcal{Z}_{k,\text{keep}}^{'1} | \mathcal{Z}_{k,\text{keep}}^{'0}\mathcal{Z}_{k,\text{keep}}^{'\text{rest}}E_{k}) 
&\geq& n_{k,1} - H_{\text{max}}^{\epsilon''_{k}}(\mathcal{X}_{k}^{1} | \mathcal{X}_{k}'^{1})\nonumber \\ 
&\geq& n_{k,1} - n_{k,1} h(e_{k,1})\nonumber.
\end{eqnarray}
Using the above equation in equation~\eqref{eq:chain 1},  we get
\begin{eqnarray} \label{eq:chain 3}
H_{\text{min}}^{\varepsilon_{k}}(\mathcal{Z}'_{k,\text{keep}} | E_{k}) &\geq& n_{k,0} + n_{k,1}\left[1 - h(e_{k,1})\right]- 2\log_{2}\frac{2}{\epsilon'_{k}\hat{\epsilon}_{k}}\nonumber. \\
\end{eqnarray}
We are interested in a regime where the first two terms on the RHS of equation~\eqref{eq:chain 3} are
much larger than the $\log_{2}$ term, as $\varepsilon'_k$ and $\hat{\varepsilon}_k$ are typically of the order say $10^{-5} -10^{-10}$. Therefore, if we neglect this $\log_{2}$ term, we obtain equation~\eqref{eq:mains} of the main paper,
\begin{eqnarray} \label{eq:chain 4}
H_{\text{min}}^{\varepsilon_{k}}(\mathcal{Z}'_{k,\text{keep}} | E_{k}) &\gtrapprox& n_{k,0} + n_{k,1}\left[1 - h(e_{k,1})\right].
\end{eqnarray}

 \section{Security against repudiation}\label{App:AppendixC}

We follow the approach in~\cite{Wallden2015}. If Alice tries to repudiate a message, she sends a declaration $(m, Sig_m)$ which Bob will accept and Charlie will reject. For this to happen, Bob must accept both the elements that Alice sent directly to him, and the elements that Charlie forwarded to him. In order for Charlie to reject he needs only to reject either the elements he received from Alice, or the elements Bob forwarded to him (or both). Intuitively, security against repudiation follows because of the symmetrisation performed by Bob and Charlie using the secret classical channel. In the distribution stage, to send the future message $m$, Alice uses the MDI-KGP with Bob and Charlie to generate strings of length $n_{k}= L$. Suppose that Bob holds the string $(b_1, ..., b_L)$ and  Charlie holds the string $(c_1, ..., c_L)$. 
Now, for simplicity, we consider that Alice has full power and we assume that later on, in the messaging stage, she is able to fully control the number of mismatches her signature declaration contains with $(b_1, ..., b_L)$ and $(c_1, ..., c_L)$. Let us denote the mismatch rates by $e_B$ and $e_C$ respectively. Then, the symmetrisation process means that Bob and Charlie will randomly (and unknown to Alice) receive $L/2$ elements of the other's string. We aim to show that any choice of $e_C$ and $e_B$ leads to an exponentially decaying probability of repudiation. 
Then we have two cases as in~\cite{Wallden2015}:\\
\\Case 1: First, let us assume that $e_C > s_a$. In this case, Bob receives $L/2$ elements from the set $\{c_1, ..., c_L\}$, which contains exactly $e_CL$ mismatches with Alice's future declaration. In order to accept the message, Bob must get fewer than $s_aL/2$ errors. Using \cite{Hoeffding1963} we can bound the probability that Bob gets fewer than $s_aL/2$ mismatches as
\begin{equation}
\begin{split}
&Pr(\text{Bob gets less than $s_aL/2$ mismatches from Charlie})\\
& \leq \exp[-(e_C-s_a)^2L].
\end{split}
\end{equation}
To repudiate, Alice must make Bob accept the message, which means that Bob must accept both the part received from Alice and the part received from Charlie. Since $Pr(A \cap B) \leq \min\{Pr(A), Pr(B)\}$ the probability of repudiation must be less than or equal to the above expression, and so must also decrease exponentially.\\
\\Case 2: Suppose $e_C \leq s_a$. In this case, if $e_B>s_a$, the above argument shows that it is highly likely that Bob will reject the message, so we examine only the case where $e_B \leq s_a$. Consider first the set $\{b_1, ..., b_L\}$. We can use the same arguments as above to bound the probability of selecting more than $s_vL/2$ mismatches as
\begin{equation}
\begin{split}
&Pr(\text{Charlie gets more than $s_vL/2$ mismatches from Bob}) \\
&\leq \exp[-(s_v-e_B)^2L].
\end{split}
\end{equation}
Then, Alice succeeds if Charlie finds more than $s_vL/2$ mismatches either from the set $\{b_1, ..., b_L\}$ or the set $\{c_1, ..., c_L\}$. Using $Pr(A \cup B) \leq Pr(A) + Pr(B)$, we can see that, for the choice of $e_B, e_C \leq s_a$, we have
\begin{equation}
\begin{split}
& Pr(\text{Charlie gets more than $s_vL/2$ mismatches}) \\
& \leq 2\exp[-(s_v-s_a)^2L].
\end{split}
\end{equation}
So again, the probability of Alice successfully repudiating decreases exponentially in the size of the signature, and Alice's best strategy would be to pick $e_B = e_C = \frac{1}{2}(s_v+s_a)$, in which case
\begin{equation}
Pr(\text{Repudiation}) \leq 2\exp\left[-\frac{1}{4}(s_v-s_a)^2L\right].
\end{equation}

\section{Calculation of the number of quantum transmissions required per signed bit}
\label{App:AppendixD}

\subsection{Parameters and constraints}
Similar to~\cite{Amiri2015}, the correctness and security of the protocol depends on the three equations \eqref{eq:honab}, \eqref{eq:rep} and \eqref{eq:forge}, which in turn depend on the choice of parameters $s_a$ and $s_v$. The parameters are considered such that $\overline{E}_{k} < s_a < s_v < p_{E}$. We say that $\overline{E}_{k}$ is the maximum of the worst-case error rates that Alice makes with Bob's key (found from the Alice-Bob MDI-KGP), and the worst-case error rates Alice makes with Charlie's key (found from the Alice-Charlie MDI-KGP). Similarly, $p_{E}$ is the minimum of the adversary's error rates found from the Alice-Bob and Alice-Charlie MDI-KGP. We follow \cite{Amiri2015} to choose the parameters that minimise the number of quantum transmissions required per signed bit. This will be larger than the signature length, $L$, due to factors such as channel loss, detection efficiency and parameter estimation procedures. Because of this, Bob will have to transmit more than $L$ quantum states to generate a signature of length $L$.

In the next section, we will calculate the length of the signature and the number of quantum transmissions
necessary to sign a message with a security level of $10^{-5}$. This means that the probabilities of honest abort, forging and repudiation, given respectively by \eqref{eq:honab}, \eqref{eq:forge} and \eqref{eq:rep}, are all less than $10^{-5}$. To find the length per possible one-bit message, of the signature necessary to securely sign a one-bit message, we must first choose the parameters $s_a$ and $s_v$. That is, a signature sequence of length $L$ needs to be transmitted for the possible message ``0", and for the possible message `1", so that the total signature sequence has length $2L$. Ideally, our choice would minimise $L$. We choose to set $\varepsilon_{PE} = 10^{-5}$ and
\beq
s_a = \overline{E}_{k} + \frac{p_{E} - \overline{E}_{k}}{3}, \:\:\:\:\:\: s_v = \overline{E}_{k} + \frac{2(p_{E} - \overline{E}_{k})}{3}.
\eeq
These may not be the optimal choices of these parameters. However, a natural choice would be to choose the parameters in order to equally partition the gap between $\overline{E}_{k}$ and $p_{E}$. 

\subsection{The number of quantum transmissions required per signed bit}

In this section, we use experimental data provided by~\cite{Ursin2007} to give an optimal estimate of the number of states Bob needs to transmit over a $50$ km quantum channel to securely sign a one bit message. We set $\epsilon_{PE} = 10^{-5}$ in all equations that follow. The experiment in \cite{Ursin2007} considers a free-space channel, we assume a fibre-based channel with a loss coefficient of $0.2$ dB/km. 
Here, we consider standard single-photon detectors where the detection efficiency of the relay is 
$14.5\%$ and the background rate is $6.02\times10^{-6}$. The overall misalignment in the channel is assumed to be $1\%$ and the bound is fixed to be $\varepsilon_k = 10^{-10}$. The other parameters involved are:
\begin{itemize}
\item Source: 1 GHz pulse rate
\item Basis probabilities: $p_Z = 62.5\%$, $p_X=37.5\%$.
\item Intensity levels: $(s, d_1, d_2) = (0.18, 0.09, 5\times10^{-4})$.
\item Intensity probabilities: $p_s = 50\%$, $p_{d1} = p_{d2} = 25\%$.
\end{itemize}

We consider the total number of signals sent by Bob to be $5.58\times10^{12}$, and find the raw key to contain $9.42\times 10^{6}$ bit values from $Z$ basis measurement outcomes. Assuming that $5.5\%$ of the detected signals are used for error rate estimation ($R_k = 5.18\times10^{5}$), we obtain a signature length of $n_k = 8.9\times 10^{6}$. Of these, Bob will randomly choose $n_k/2 = 4.45\times 10^{6}$ to be $\mathcal{Z'}_{k,\text{keep}}$, another $n_k/2$ will be used as $\mathcal{Z'}_{k,\text{forward}}$.

For the given intensity levels and intensity choice probabilities, we observe an error rate in the $Z$ basis given by $E_{k}^{a_{s},b_{s}} = 2.07\%$. This error rate arises from the channel misalignment together with the dark-count rate of the detectors. We can then use Eq.\eqref{eq:actualerrorrate} to upper bound the true error rate as $\overline{E}_{k}^{a_{s},b_{s}} = 2.39\%$.

We use Appendix \ref{App:AppendixA} to estimate the relevant parameters by setting all $\varepsilon$ as $10^{-10}$, and thereby we can calculate the min-entropy. Finally, setting $\varepsilon_k = 10^{-10}$, we get
\beq \label{eq:num_min_ent}
H_{\text{min}}^{\varepsilon_{k}}(\mathcal{Z'}_{k,\text{keep}} | E_{k}) = 8.69 \times 10^{5}.
\eeq
Then using~\eqref{eq:EveError} we find $p_E$ as $3.02\%$, and so we obtain $s_a = 2.60\%$ and $s_v=2.81\%$. Setting $g$ as $10^{-5}$ and substituting these values into equations \eqref{eq:honab}, \eqref{eq:forge} and \eqref{eq:rep}, we find $Pr(\text{Honest Abort}) = 2.00\times 10^{-5}$, $Pr(\text{Forge}) = 3\times 10^{-5}$, and $Pr(\text{Repudiation}) = 9.857\times 10^{-5}$. Thus we observe that when $5.58\times10^{12}$ states are transmitted, the protocol is secure to a level of the order of $10^{-5}$ for a distance of $50$ km. The analysis for the other cases shown in Tables II and III is done in a similar way.

\bibliographystyle{apsrev4-1}

\begin{thebibliography}{10}
\bibitem{footnote1} If we exceed the limits of $M_{k}^{a,b}$ and $N_{k}^{a,b}$, we then select at random $M_{k}^{a,b}$ and $N_{k}^{a,b}$ elements for the analysis.
\end{thebibliography}

\end{document}